\documentclass[twocolumn,conference]{IEEEtran}
\usepackage[T1]{fontenc}
\usepackage{amsthm}
\usepackage{graphicx}

\makeatletter
\theoremstyle{plain}

\theoremstyle{plain}
\newtheorem{lem}{\protect\lemmaname}

\IEEEoverridecommandlockouts
\usepackage{ifpdf}
\usepackage{cite}
\ifCLASSOPTIONcompsoc
\usepackage[caption=false,font=normalsize,labelfont=sf,textfont=sf]{subfig}
\else
\usepackage[caption=false,font=footnotesize]{subfig}
\fi
\usepackage{amssymb}
\usepackage{amsmath}
\usepackage{float}
\usepackage{array}
\usepackage{makecell}
\newenvironment{sequation}{\begin{equation}\small}{\end{equation}}

\makeatother

\providecommand{\lemmaname}{Lemma}
\providecommand{\theoremname}{Theorem}

\begin{document}

\title{A Novel Scheme for Downlink Opportunistic Interference Alignment}

\author{Haijing Liu, Hui Gao, Wei Long, Tiejun~Lv\\Key Laboratory of Trustworthy
Distributed Computing and Service, Ministry of Education\\School
of Information and Communication Engineering\\Beijing University
of Posts and Telecommunications, Beijing, China 100876\\Email: \{Haijing\_LIU,
huigao, longwei, lvtiejun\}@bupt.edu.cn%
\thanks{This work is financially supported by the National Natural Science
Foundation of China (NSFC) under Grant No. 61271188.%
}}
\maketitle
\begin{abstract}
In this paper we propose a downlink codebook-based opportunistic interference
alignment (OIA) in a three-cell MIMO system. A codebook composed of
multiple transmit vector sets is utilized to improve the multiuser
selection diversity. The sum rate increases as the size of the codebook
grows. In addition, during the user selection, effective channel gain
and alignment metric are combined to generate a novel criterion, which
improves the system performance, especially at low SNR. Furthermore,
a threshold-based feedback approach is introduced to reduce the feedback
load in the proposed scheme. Both the analytical results and simulations
show that the proposed scheme provides significant improvement in
terms of sum rates with no feedback load growth and slight increase
of complexity.
\end{abstract}

\section{Introduction}

\setlength{\textfloatsep}{8pt plus 2pt minus 4pt}
\setlength\abovedisplayskip{8pt}
\setlength\belowdisplayskip{6pt}

With the exponential growth in mobile data traffic, interference has
been one of the major challenges in wireless communication. Interference
alignment (IA) \cite{cadambe_interference_2008} is a technique recently
introduced to improve the performance of interference networks. Unfortunately,
extensive channel station information and a large amount of computation
is required to achieve the optimal DoFs \cite{peters_cooperative_2011},
which makes IA too complicated to be implemented in practice. Motivated
by opportunistic beamforming (OBF) \cite{viswanath_opportunistic_2002},
OIA schemes are developed in \cite{lee_opportunistic_2010,lee_achievable_2013,leithon_new_2012,gao_new_2013,gao_limited_2014,chen_performance_2014},
which only require limited feedback and modest computational complexity.
Though OIA takes advantage of multiuser diversity via opportunistic
user equipment (UE) scheduling, \cite{lee_achievable_2013} proves
that the number of required UEs grows with an exponential scale in
order to achieve an optimal DoF. In practical systems, the number
of UEs is usually limited, so the improvement of sum rate performance
via OIA is not obvious. On the other hand, UEs are selected from the
perspective of interference reduction in OIA, while the selection
is done from the point of view of channel gains in Maximum SNR (MAX-SNR)
scheduling. OIA outperforms MAX-SNR in an interference limited environment
while MAX-SNR provides better performance in a noise limited environment.
Neither of them have a wide SNR range of application.

In this paper, we propose a downlink codebook-based opportunistic
interference alignment (COIA) scheme. Compared with the conventional
OIA schemes,three improvements are made:

($i$) A codebook composed of multiple transmit beamforming vector
sets for three base stations (BSs) is utilized to bring more selection
diversity. More specifically, besides the UE scheduling, BSs select
a transmit beamforming vector set from the codebook to enhance system
sum rate. Consequently, fewer UEs are needed in COIA to achieve the
same sum rate compared with the conventional OIA. Particularly, with
the theoretical analysis of the expectation of the alignment metric
value, we can pre-calculate the required numbers of candidate UEs
with various codebook sizes for the same sum rate as that of the conventional
OIA. Note that code-book based uplink OIA schemes have recently been
proposed in \cite{yang_codebook-based_2013,lee_codebook-based_2014}.
Our downlink COIA is completely different from them because the codebook
in our scheme is utilized to exploit the selection diversity, while
the codebook in \cite{yang_codebook-based_2013,lee_codebook-based_2014}
is used to reduce the feedforward load. We propose our downlink COIA
scheme to improve the sum rate performance, while they focus on the
UE and feedback bit scaling law with their uplink COIA schemes.

($ii$) An effective UE selection metric adaptively balancing the
noise and interference power is introduced to overcome the shortcoming
of OIA schemes at low SNR. With the above two improvements, COIA achieves
better sum rate performance than MAX-SNR scheduling and the conventional
OIA the same number of candidate UEs.

($iii$) When the codebook size is very large, the feedback load becomes
unacceptable of our previous COIA scheme in \cite{liu_codebook-based_2013}.
In this paper, a threshold-based feedback scheme, which has never
been discussed in OIA to the best of our knowledge, is explored to
reduce the system feedback load in our COIA. We address the relationship
between the threshold value and feedback load by an explicit expression,
so that the feedback load of the proposed scheme can be adjusted to
the same as that of the conventional OIA by setting an appropriate
threshold.

Throughout the paper, we describe matrices and vectors by bold upper
and lower case letters. $\mathbf{A}^H$, $\lambda_{\textnormal{max}}(\mathbf{A})$,
$\mathbf{v}_{\textnormal{max}}(\mathbf{A})$, $\lVert{\mathbf{A}}\rVert$
and $\mathbf{A}^{-1}$ denote the conjugate transpose, the largest
eigenvalue, the eigenvector corresponding to the largest eigenvalue,
$L_2$-norm and the inverse of matrix $\mathbf{A}$, respectively.

\section{System Model }

We consider a 3-cell MIMO downlink system with a single BS and $K$
UEs in each cell. Both the BS and the UEs are each equipped with two
antennas. In the $i$-th cell, $i=1,2,3,$ the BS sends a data stream
to a scheduled UE with a normalized transmit beamforming vector $\mathbf{w}_i$,
where $\lVert{\mathbf{w}_i}\rVert=1$. For convenience we denote the
$k$-th UE in the $i$-th cell as UE $[k,i]$, where $1\leq k\leq K,k\in\mathbb{N}$
and $i=1,2,3$. Quasi-static channels between BSs and UEs are assumed.
The received signal at UE $[k,i]$ is \begin{sequation}\label{eq1}\mathbf{y}_{k_{i}}=\sqrt{P_S}\mathbf{H}_{k_{i},i}\mathbf{w}_ix_i+\sqrt{P_I}\sum_{j=1,j\not=i}^{3}\mathbf{H}_{k_{i},j}\mathbf{w}_{j}x_{j}+\mathbf{n}_{k_{i}},\end{sequation}where
$\mathbf{H}_{k_{i},j}\in\mathbb{C}^{2\times2}$ is the channel matrix
from the BS in the $j$-th cell to UE $[k,i]$. Elements of $\mathbf{H}_{k_{i},j}$
are independent and identically distributed (i.i.d.) circularly symmetric
complex Gaussian random variables with zero mean and unit variance.
$x_j$ is the signal transmitted by the $j$-th BS with a transmit
power constraint $\mathbb{E}[|x_j|^2]=1$. $\mathbf{n}_{k_{i}}\in\mathbb{C}^{2\times1}\sim\mathcal{CN}(\mathbf{0},\sigma^2_n\mathbf{I})$
is the additive complex Gaussian noise at UE $[k,i]$. $P_S$ stands
for the received data power and $P_I$ is the received average interference
power from each interfering BS. Denoting the receive beamforming vector
of UE $[k,i]$ by $\mathbf{v}_{k_{i}}\in\mathbb{C}^{2\times1}$, the
received signal after receive beamforming is \begin{sequation}\label{eq2}\mathbf{v}^H_{k_{i}}\mathbf{y}_{k_{i}}\!=\!\sqrt{P_S}\mathbf{v}^H_{k_{i}}\mathbf{H}_{k_{i},i}\mathbf{w}_ix_i+\sqrt{P_I}\!\sum_{j=1,j\not=i}^{3}\!\mathbf{v}^H_{k_{i}}\mathbf{H}_{k_{i},j}\mathbf{w}_{j}x_{j}+\mathbf{v}^H_{k_{i}}\mathbf{n}_{k_{i}}.\end{sequation}We
also assume there exist low-rate but reliable and delay-free backhaul
links between each UE with its relevant BS as well as among the BSs.

Based on \eqref{eq2}, the signal-to-interference-plus-noise ratio
(SINR) of the data stream of UE $[k,i]$ is given by \begin{sequation}\label{eq3}\textnormal{SINR}_{k_{i}}=\frac{P_S|\mathbf{v}^H_{k_{i}}\mathbf{H}_{k_i,i}\mathbf{w}_i|^2}{\sigma^2_n+P_I\sum_{j=1,j\not=i}^{3}|\mathbf{v}^H_{k_{i}}\mathbf{H}_{k_{i},j}\mathbf{w}_{j}|^2}.\end{sequation}

\section{Conventional Opportunistic User Selection Schemes}

In both OBF and conventional OIA, the transmit beamforming vectors
$\mathbf{w}$ are all generated randomly, while the UE selection criteria
are significantly different. We discuss three opportunistic UE selection
schemes (i.e., MAX-SINR, MAX-SNR and the conventional OIA) in this
section.

\subsection{MAX-SINR}

MAX-SINR has been shown to be an optimal opportunistic UE selection
scheme in the sense of sum rate so far\cite{lee_achievable_2013}.
The receive beamforming vector of UE $[k,i]$ is $\mathbf{v}^{\textnormal{MAX-SINR}}_{k_{i}}=\mathbf{v}_{\textnormal{max}}(\mathbf{A}_{k_{i}}^{-1}\mathbf{B}_{k_{i}})$
to maximize $\textnormal{SINR}_{k_{i}}$, where $\mathbf{A}_{k_{i}}=\sigma^2_n\mathbf{I}+P_I\sum_{j=1,j\not=i}^{3}\mathbf{H}_{k_{i},j}\mathbf{w}_j\mathbf{w}^H_j\mathbf{H}^H_{k_{i},j}$
and $\mathbf{B}_{k_{i}}=P_S\mathbf{H}_{k_{i},i}\mathbf{w}_i\mathbf{w}^H_i\mathbf{H}^H_{k_{i},i}$.
The corresponding SINR is \begin{sequation}\textnormal{SINR}_{k_{i}}=\lambda_{\textnormal{max}}(\mathbf{A}_{k_{i}}^{-1}\mathbf{B}_{k_{i}}).\nonumber\end{sequation}
The UE with the largest SINR is selected, i.e., \begin{sequation}\label{eq7}k^{\textnormal{MAX-SINR}}_{i}=\mathop{\arg\max}_{1{\leq}k_{i}{\leq}K}\textnormal{SINR}_{k_{i}}.\nonumber\end{sequation}

\subsection{MAX-SNR}

In MAX-SNR, the receive beamforming vector of UE $[k,i]$ is designed
as $\mathbf{v}^{\textnormal{MAX-SNR}}_{k_{i}}=\frac{\mathbf{H}_{k_{i},i}\mathbf{w}_i}{\lVert\mathbf{H}_{k_{i},i}\mathbf{w}_i\rVert}$
to maximize the SNR. The corresponding SNR is $\textnormal{SNR}_{k_{i}}=\frac{P_S\lVert\mathbf{H}_{k_{i},i}\mathbf{w}_i\rVert^2}{\sigma^2_n}.$
In homogeneous network, each UE calculates its effective channel gain
\begin{sequation}\label{eq5}\beta_{k_{i}}=\lVert\mathbf{H}_{k_{i},i}\mathbf{w}_i\rVert^2\end{sequation}
and informs the corresponding BS. The BS selects the UE with the largest
SNR, i.e., \begin{sequation}k^{\textnormal{MAX-SNR}}_{i}=\mathop{\arg\max}_{1{\leq}k_{i}{\leq}K}\beta_{k_{i}}.\end{sequation}

\subsection{Conventional OIA}

In OIA\cite{lee_achievable_2013}, the UE whose interference signals
are most aligned with each other is selected. The alignment of interfering
signals is measured by their chordal distance. The metric value of
UE $[k,i]$ is \begin{sequation}\label{eq8}\gamma_{k_{i}}=\frac{\lVert\mathbf{w}^H_{i'}\mathbf{H}^H_{k_{i},i'}\mathbf{H}_{k_{i},i''}\mathbf{w}_{i''}\rVert^2}{\lVert\mathbf{H}_{k_{i},i'}\mathbf{w}_{i'}\rVert^2\cdot\lVert\mathbf{H}_{k_{i},i''}\mathbf{w}_{i''}\rVert^2},\end{sequation}where
$i'$ is the $i$-th element of vector $[2,3,1]$, and $i''$ is the
$i$-th element of vector $[3,1,2]$. Each UE sends the value back
to the relevant BS. The preferred UE in the $i$th cell is\begin{sequation}\label{eq9}k^{\textnormal{OIA}}_{i}=\mathop{\arg\max}_{1{\leq}k_{i}{\leq}K}\gamma_{k_{i}}.\end{sequation}

\section{Novel Codebook-Based OIA Scheme}

In this section, we propose an OIA scheme with a codebook of transmit
beamforming vector sets. A novel selection criterion adaptive to noise
and interference power as well as a threshold-based feedback are further
developed to enhance the sum rate performance and control the feedback
load of the proposed scheme.

\subsection{Codebook-Based OIA\label{sub:Codebook-Based-OIA}}

In codebook-based downlink OIA, BSs choose transmit beamforming vectors
from multiple vectors in a codebook every time slot. The codebook
composed of transmit beamforming vector sets is denoted by $\mathcal{C}=\{\mathbf{c}_1,\dots,\mathbf{c}_S\},$
where $\mathbf{c}_s$ is the concatenation of the $s$-th set of random
unit-norm transmit beamforming vectors, i.e., $\mathbf{c}_s=[\mathbf{w}^H_{1,s},\mathbf{w}^H_{2,s},\mathbf{w}^H_{3,s}]^H\in\mathbb{C}^{6\times1}$,
and $S$ is the size of the codebook. All the UEs and BSs know the
codebook $\mathcal{C}$.

The UE selection and data transmission in COIA is shown as follows:

\textbf{\emph{Step 1}}: Each BS broadcasts pilots for channel estimation.
Every UE obtains channel estimations $\hat{\mathbf{H}}_{k_{i},i}$
and $\hat{\mathbf{H}}_{k_{i},j}$.

\textbf{\emph{Step 2}}: Using the estimated channel information,
each UE calculates $S$ alignment metric values for $S$ transmit
beamforming vector sets in $\mathcal{C}$. The alignment metric value
of UE $[k,i]$ for the $s$-th transmit beamforming vector set is
\begin{sequation}\label{eq11}\gamma_{k_{i},s}=\frac{\lVert\mathbf{w}^H_{i',s}\hat{\mathbf{H}}^H_{k_{i},i'}\hat{\mathbf{H}}_{k_{i},i''}\mathbf{w}_{i'',s}\rVert^2}{\lVert\hat{\mathbf{H}}_{k_{i},i'}\mathbf{w}_{i',s}\rVert^2\cdot\lVert\hat{\mathbf{H}}_{k_{i},i''}\mathbf{w}_{i'',s}\rVert^2}.\end{sequation}Each
UE feeds the analog metric values back to the BS in its own cell.

\textbf{\emph{Step 3}}: BSs exchange the analog feedback, then select
the preferred transmit beamforming vector set as well as the corresponding
served UEs. Regarding a specific transmit beamforming vector set $\mathbf{c}_s$,
we first find the UE with the largest alignment metric value in the
$i$-th cell and the corresponding metric value, denoted by \begin{sequation}\label{eq12}\bar{k}_{i,s}=\mathop{\arg\max}_{1{\leq}k_{i}{\leq}K}\gamma_{k_{i},s},\quad \bar{\gamma}_{i,s}=\max_{1{\leq}k_{i}{\leq}K}{\gamma_{k_{i},s}}.\end{sequation}After
that, we calculate the average of the largest alignment metric values
of three cells for $\mathbf{c}_s$, which is given by \begin{sequation}\label{eq30}\bar{\gamma}_{s}=\frac{1}{3}\sum_{i=1}^3\bar{\gamma}_{i,s}.\end{sequation}The
preferred transmit beamforming vector set is then selected among all
sets in the codebook as \begin{sequation}\label{eq13}s^*=\mathop{\arg\max}_{1{\leq}s{\leq}S}\bar{\gamma}_{s},\end{sequation}which
means we choose the codeword to maximize the average of the largest
alignment metric values of three cells. Once $s^*$ is determined,
the selected transmit beamforming vector for the $i$-th transmitter
is $\mathbf{w}_{i,s^*}$ and the preferred UE being served in the
$i$-th cell is $\bar{k}_{i,s^*}$.

\textbf{\emph{Step 4}}: Each BS serves the selected UE with the preferred
transmit beamforming vector.

\subsection{Analysis of Codebook-Based OIA\label{sub:Analysis-of-Codebook-Based}}

We provide a theoretical analysis of the codebook-based OIA. The expectation
of the alignment metric value of the selected UE increases as $S$
grows. In other words, compared with the conventional OIA ($S=1$),
the interfering signals of the selected UE are aligned more and more
closely when the codebook size increases in COIA.

The expectation of $\bar{\gamma}_{s^*}$ (i.e., the average of alignment
metric value of the selected UE $\bar{k}_{i,s^*}$), is approximately
given by \begin{sequation}\label{eq15}\begin{split}&\mathbb{E}[\bar{\gamma}_{s^*}]\\&=\frac{S}{(\mathbb{B}(a,b))^S}{\!}\cdot{\!} P(a,b,S){\!}\cdot{\!} \mathbb{B}(aS+k_1+\cdots+k_{S-1}+1,b),\end{split}\end{sequation}where
$\mathbb{B}(\cdot)$ is the beta function, $a=\frac{3K(K+2)}{K+1}$,
$b=\frac{3(K+2)}{K+1}$, and \begin{sequation}\begin{split}&P(a,b,S){\!}\\&={\!}\sum_{k_1=0}^\infty{\!}\cdots{\!}\sum_{k_{S-1}=0}^\infty{\!}\frac{(1-b)_{k_1}{\!}\cdots{\!} (1-b)_{k_{S-1}}}{(a{\!}+{\!}k_1){\!}\cdots {\!}(a{\!}+{\!}k_{S-1})k_1!{\!}\cdots{\!} k_{S-1}!}.\end{split}\end{sequation}
See Appendix for the derivation of \eqref{eq15}.

\begin{table}[t] \caption{$\mathbb{E}[\bar{\gamma}_{s^*}]$ of various $K$ and $S$.} \label{tab:t1}
\centering
\begin{tabular}{c|c|c|c|c}
\hline
\makecell*[c] {} & $S=1$ & $S=2$ & $S=3$ & $S=4$
\tabularnewline
\hline
\hline
\makecell*[c]{$K=10$} & 0.9091 & 0.9351 & 0.9461 & 0.9529
\tabularnewline
\hline
\makecell*[c]{$K=15$} & 0.9375 & 0.9559 & 0.9635 & 0.9680
\tabularnewline
\hline
\makecell*[c]{$K=20$} & 0.9524 & 0.9666 & 0.9725 & 0.9759
\tabularnewline
\hline
\end{tabular}
\end{table}

Given number of UEs $K$ and codebook size $S$, we can get the expectation
efficiently because $\mathbb{B}(\cdot)$ can be calculated directly
in MATLAB. TABLE \ref{tab:t1} shows $\mathbb{E}[\bar{\gamma}_{s^*}]$
for various $K$ and $S$. It can be seen that the expectation increases
with the growth of $S$ for the same $K$.

Since the average rate of the selected UE increases as the expectation
of its alignment metric value grows \cite{lee_achievable_2013}, with
the help of \eqref{eq15}, we can get the number of UEs $K$ with
variable codebook sizes $S$ for the same expectation, i.e., the same
sum rate performance. For example, when $K=20,S=1$, $\mathbb{E}[\bar{\gamma}_{s^*}]=\textnormal{0.9524}$.
Then letting $S=4$ and setting the left hand side value of $\eqref{eq15}$
as 0.9524, we can get the required number of UEs $K=10$ for the same
performance. Only half of UEs are needed in COIA with $S=4$ codebook
compared with the conventional OIA.

\subsection{Hybrid Criterion in COIA}

The effective channel gain in MAX-SNR of UE $[k,i]$ with the $s$-th
transmit beamforming vector is defined as \begin{sequation}\label{eq18}\beta_{k_{i},s}=\lVert\hat{\mathbf{H}}_{k_{i},i}\mathbf{w}_{i,s}\rVert^2.\end{sequation}
We introduce a hybrid criterion with \eqref{eq11} and \eqref{eq18},
which is given by \begin{sequation}\label{eq19}\alpha_{k_{i},s}=[0,(1-\theta)]^+\cdot\gamma_{k_{i},s}+\theta\cdot \beta_{k_{i},s},\end{sequation}
where $\theta=\frac{P_S/P_I}{P_S/\sigma_n^2}=\frac{\sigma_n^2}{P_I}$
and $[x,0]^+=\max(x,0).$ The BSs select the transmit beamforming
vector set and UEs in the same way as that mentioned in Part. \ref{sub:Codebook-Based-OIA},
except replacing $\gamma_{k_{i},s}$ with $\alpha_{k_{i},s}$ in \eqref{eq12}.
We can see that when the power of interference is smaller than that
of noise, i.e., the system is at low SNR, the hybrid metric value
only depends on the effective channel gain. With the increase of interference
power, the proportion of the effective channel gain decreases and
the effect of OIA UE selection is enhanced. At very high SNR, the
hybrid metric value is almost equal to the OIA metric value. With
the proposed hybrid criterion adaptive to noise and interference power,
the COIA scheme achieves better sum rate performance in both low and
high SNR regions.

\subsection{Threshold-Based Feedback in COIA}

In the OIA scheme proposed in \cite{lee_achievable_2013}, every UE
feeds back an alignment metric value to the corresponding BS, which
we refer to as full feedback. $K$ values are needed to complete a
UE selection in each cell. In COIA, if full feedback is adopted, the
amount of feedback will be $K\cdot S$ due to the utilization of the
$S$ size transmit beamforming vector codebook. The feedback load
becomes unacceptable when $S$ is large. Here we propose a threshold-based
feedback technique to reduce the feedback needs (by more than 75\%)
while preserving the sum rate performance in COIA. Similar techniques
are introduced in \cite{gesbert_how_2004}, \cite{vegard_exploiting_2005}.
However, they take only signal and noise into consideration and ignore
interference, which degrades their performance in multi-cell systems.

In the proposed threshold-based feedback scheme, each UE compares
its selection metric value to a predefined threshold $T$ and decides
locally whether it sends feedback to the BS, only those who fall above
$T$ are allowed to be fed back. BSs make selections with the feedback.
If no feedback is received by all three BSs, transmit beamforming
vector set and UE in each cell is selected randomly.

Choosing a proper threshold is critical. We first characterize the
statistics of the alignment metric $\gamma$ and the effective channel
gain $\beta$ in terms of cumulative distributive function (CDF) and
probability density function (PDF). As \eqref{eq8} shows, the alignment
metric $\gamma$ is related to the chordal distance between two vectors.
Using the results of \cite{au-yeung_performance_2007},the CDF of
$\gamma$, denoted by $F_{\gamma}(x)$ is given by \begin{equation}\label{eq28}F_{\gamma}(x)=P(\gamma\leq x)=x,\quad 0\leq x\leq 1.\end{equation}The
PDF of $\gamma$ is \begin{equation}\label{eq23}f_{\gamma}(x)=1,\quad 0\leq x\leq 1.\end{equation}
The effective channel gain $\beta$ is defined as \eqref{eq5}. With
a certain unit-norm vector $\mathbf{w}$, $\beta$ has a central chi-square
distribution. The CDF of $\beta$ is given by \begin{equation}\label{eq29}F_{\beta}(y)=P(\beta\leq y)=1-e^{-y}(1+y),\quad y\geq 0.\end{equation}
The PDF of $\beta$ is \begin{equation}\label{eq24}f_{\beta}(y)=ye^{-y},\quad y\geq 0.\end{equation}The
normalized average feedback load $\bar{F}$ is defined as the ratio
of the average load per selection to the total amount of full feedback
($KS$) in each cell. Apparently, with a threshold $T$, we have \begin{equation}\label{eq20}\bar{F}^{\textnormal{OIA}}(T)=1-F_{\gamma}(T)\end{equation}
and \begin{equation}\label{eq21}\bar{F}^{\textnormal{MAX-SNR}}(T)=1-F_{\beta}(T).\end{equation}
For a given feedback load requirement $\bar{F}$ (e.g., 1/4), we can
get the threshold $T^{\textnormal{OIA}}$ and $T^{\textnormal{MAX-SNR}}$
with \eqref{eq20} and \eqref{eq21}, respectively.

In COIA, the metric value $\alpha$ is given by \eqref{eq19}. With
\eqref{eq23} and \eqref{eq24}, when $0<\theta<1$, the CDF of $\alpha$
is given by (we omit the derivation due to space limitations) \begin{sequation}\label{eq25}F_{\alpha}(z)=\left\{\begin{aligned}&\frac{1}{1-\theta}-\frac{e^{-\frac{z}{\theta}}(\theta+z)}{(1-\theta)\theta}, \quad 0\leq z<1-\theta \\&-\frac{e^{-\frac{z}{\theta}} \left(\theta+z-e^{\frac{1-\theta}{\theta}} (2\theta-1+z)\right)}{(1-\theta)\theta}, z\geq 1-\theta\\ \end{aligned}\right. .\end{sequation}
When $\theta\geq 1$, $\alpha=\theta\cdot\beta$, the CDF of $\alpha$
is given by \begin{sequation}\label{eq26}F_{\alpha}(z)=1-e^{-\frac{z}{\theta}}(1+\frac{z}{\theta}),\quad z\geq 0.\end{sequation}
The normalized average feedback load $\bar{F}$ is of COIA is \begin{sequation}\label{eq27}\bar{F}^{\textnormal{COIA}}(T)=1-F_{\alpha}(T).\end{sequation}We
can choose a proper threshold $T^{\textnormal{COIA}}$ with \eqref{eq27}.
It should be remarked that $T^{\textnormal{COIA}}$ is related to
$\theta$, i.e., $T^{\textnormal{COIA}}$ is adaptive to noise and
interference power, because we consider both signal channel quality
and interference condition in COIA. It is different form $T^{\textnormal{OIA}}$
and $T^{\textnormal{MAX-SNR}}$ as \eqref{eq20} and \eqref{eq21}
are only functions of $T$.

\subsection{Complexity Analysis}

We analysis the computational complexity in the UE selection step
of each UE in COIA briefly. Only complex multiplication is considered
for simplicity. Assume the number of receive antennas is $N$. For
every vector set in the codebook, the effective channel gain consumes
$\mathcal{O}(N)$ computation, and operations of $\mathcal{O}(N)$
are needed to get the alignment metric. So $2S\cdot \mathcal{O}(N)$
computation is required for each UE in COIA. Just like MAX-SNR and
the conventional OIA, the computational complexity of COIA is $\mathcal{O}(N)$.
Note that we do not take the computation of channel estimation into
account here because it is necessary for receive beamforming regardless
of UE selection schemes.

\section{Numerical Simulations}

In this section, we simulate the performance of the proposed COIA
scheme. Preferred UErs are selected with different schemes, then the
selected UE $k^*_{i}$ executes MAX-SINR receive beamforming. It should
be mentioned that we focus on the UE selection scheme while the receive
beamforming vectors design after UE selection is not studied in depth.
The sum rate is obtained according to the equation \begin{sequation}R=\sum^3_{i=1}\log_2(1+ \textnormal{SINR}_{k^*_{i}}),\end{sequation}where
$\textnormal{SINR}_{k^*_{i}}$ can be obtained by \eqref{eq3}. Perfect
channel estimation is assumed at all the UEs.

\begin{figure}[t]
\begin{centering}
\includegraphics[width=7cm]{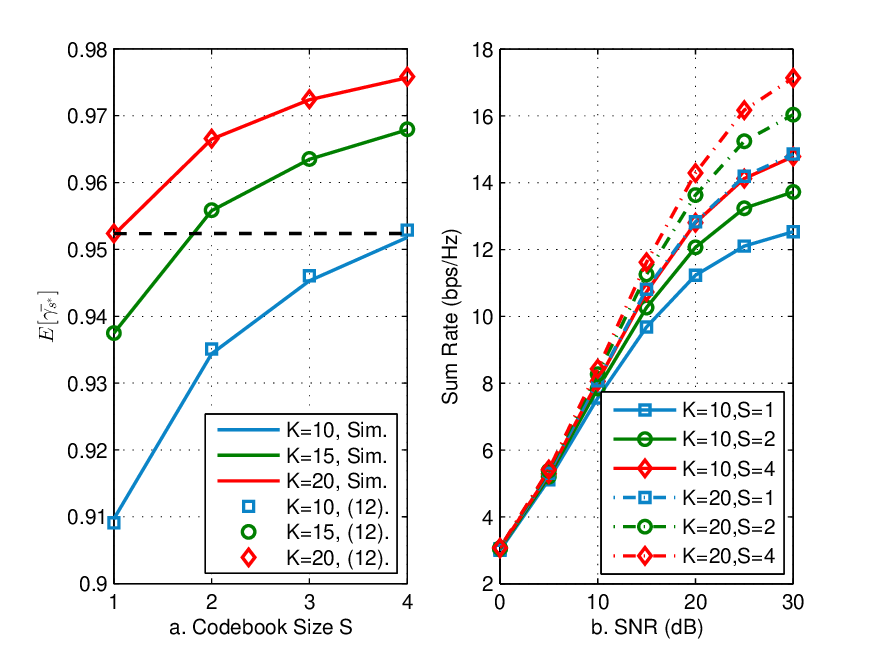}
\par\end{centering}

\caption{Expectations and sum rates of full feedback codebook-based OIA of
various $S$ and $K$. $P_S=P_I.$}
\label{fig:f12}
\end{figure}

Fig. \ref{fig:f12}. a shows the expectation of the alignment metric
value of the selected UEs in the codebook-based OIA. It is clear that
the expectation increases with the increase of codebook size $S$
for the same number of UEs $K$. Further, the configuration $K=10,S=4$
has almost the same expectation as $K=20,s=1$ has. Fig. \ref{fig:f12}.
b shows the sum rates of full feedback codebook-based OIA. The sum
rates increase with the growth of $K$ and $S$, especially at high
SNR. Only $K=10$ UEs are needed in $S=4$ codebook-based OIA to achieve
almost the same sum rate performance as $K=20$ UEs in the conventional
OIA (i.e., $S=1$), which is consistent with the analytical result
in Part. \ref{sub:Analysis-of-Codebook-Based}.

\begin{figure}[t]
\begin{centering}
\includegraphics[width=7cm]{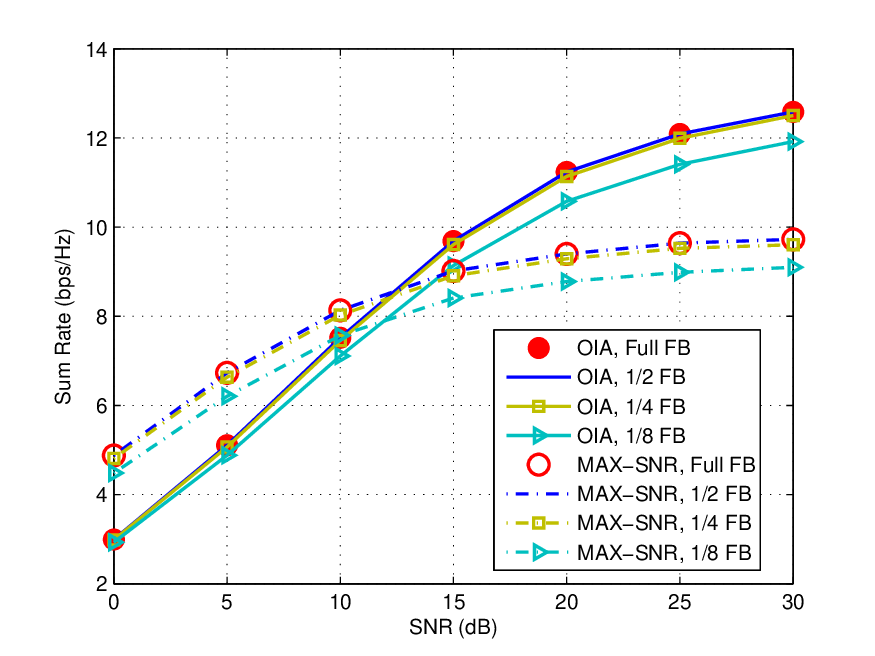}
\par\end{centering}

\caption{Sum rates of threshold-based feedback OIA and MAX-SNR. $K=10,P_S=P_I.$}
\label{fig:f21}
\end{figure}

Fig. \ref{fig:f21} shows the sum rates of threshold-based feedback
OIA and MAX-SNR with various feedback load requirment $\bar{F}$.
In OIA, the threshold value $T^{\textnormal{OIA}}$ is 0.5, 0.75 and
0.875 when $\bar{F}(T^{\textnormal{OIA}})$ is 1/2, 1/4 and 1/8, respectively.
In MAX-SNR, the threshold value $T^{\textnormal{MAX-SNR}}$ is 1.6785,
2.6925 and 3.6070 when $\bar{F}(T^{\textnormal{MAX-SNR}})$ is 1/2,
1/4 and 1/8, respectively. The sum rate loss is negligible with 1/2
and 1/4 feedback load in the threshold-based feedback scheme. It means
that a large reduction of the feedback is possible while preserving
most of the sum rate performance.

\begin{figure}[t]
\begin{centering}
\includegraphics[width=7cm]{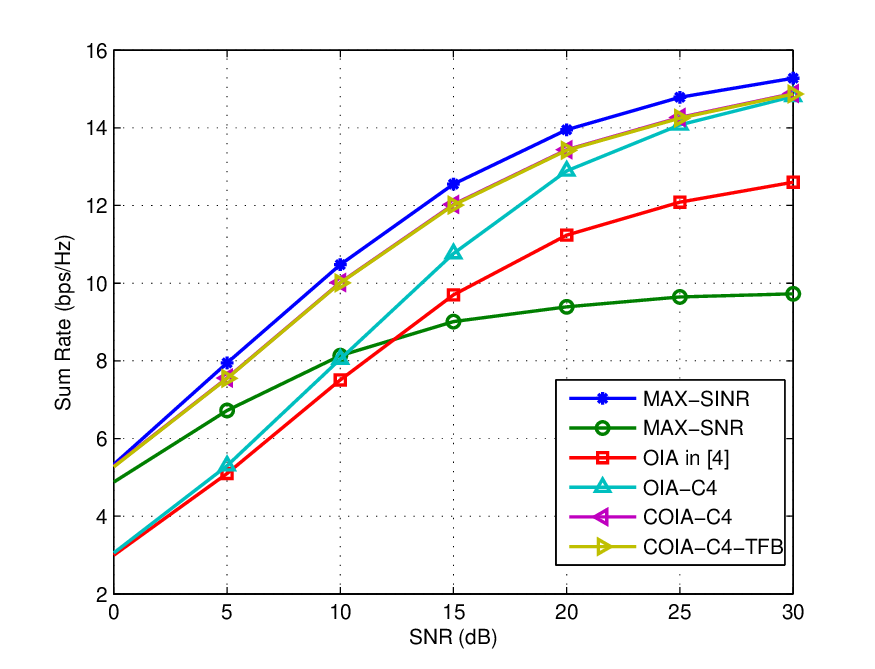}
\par\end{centering}

\centering{}\caption{Sum rates of various opportunistic UE selection schemes. $K=10,P_S=P_I.$}
\label{fig:f3}
\end{figure}

In Fig. \ref{fig:f3}, the sum rates of various schemes are shown
with $K=10$ UEs in each cell and $P_S=P_I$. The word ``C4'' in
legends means the codebook size $S=4$. The threshold-based feedback
scheme is marked as ``TFB''. For a fair comparison, we choose $\bar{F}^{\textnormal{COIA}}=1/S=1/4$
and calculate $T^{\textnormal{COIA}}$ according to \eqref{eq25},
\eqref{eq26} and \eqref{eq27}. MAX-SNR scheme outperforms the conventional
OIA \cite{lee_achievable_2013} and even codebook-based OIA in the
low SNR region but gets significant performance degradation at high
SNR. The proposed COIA with $S=4$ codebook approaches better sum
rate performance than MAX-SNR and the conventional OIA in all range
of SNR. In COIA, the sum rate performance of threshold-based feedback
is almost the same as that of full feedback, which means the proposed
COIA with threshold-based feedback outperforms MAX-SNR and the conventional
OIA with the same feedback load.

\section{Conclusions}

In this paper, we have proposed a codebook-based opportunistic interference
alignment with a hybrid selection criterion and threshold-based feedback
in a three-cell MIMO downlink system. A codebook composed of multiple
transmit vector sets is utilized to improve the multiuser selection
diversity. Effective channel gain and alignment metric are combined
to generate a novel metric for a wide SNR range of application. A
threshold is employed to reduce the feedback load in COIA. Both the
analytical results and simulations indicate that the proposed COIA
scheme provides higher sum rates in wide SNR region than the conventional
OIA scheme with the same feedback load. In the future, we will focus
on the COIA scheme with multiple data streams for each UE.

\appendix{}

Defined in \eqref{eq12}, it can be proved easily that $\bar{\gamma}_{i,s}\sim \textnormal{Beta}(K,1)$
in a similar way to \cite{lee_achievable_2013}, where $\textnormal{Beta}(\cdot)$
is the beta distribution. We omit the proof due to space limitations.
\begin{lem}
(\cite{gupta_handbook_2004}): Let $S=\sum_{i=i}^{k}X_i$ where $X_i$
are i.i.d. random variables of $Beta(\alpha,\beta)$. The distribution
of $S$ can be approximated by:\begin{sequation}Beta(e,f);e=Ff,f=\frac{F}{\sigma^2(1+F)^3}\end{sequation}
where $E=\sum\mathbb{E}[X_i]$, $F=\frac{E}{1-E}$, and $\sigma^2=\sum Var(X_i).$
\end{lem}
As $\bar{\gamma}_{i,s},i=1,2,3$ are i.i.d. beta-distributed random
variables, using Lemma 1, we can consider $\bar{\gamma}_{s}$ defined
by \eqref{eq30} as a new beta-distributed random variable, i.e.,
$\bar{\gamma}_{s}\sim \textnormal{Beta}(a,b),$ where $a=\frac{3K(K+2)}{K+1}$,
$b=\frac{3(K+2)}{K+1}.$

Let $x=\bar{\gamma}_{s^*}$ for convenience, the explicit expression
of the expectation of the maximum of i.i.d. beta-distributed random
variables $\bar{\gamma}_{1},\dots,\bar{\gamma}_{S}$ is \begin{sequation}\label{eq14}\begin{split}\mathbb{E}[\bar{\gamma}_{s^*}]&=\mathbb{E}[\max_{1\leq s\leq S}\bar{\gamma}_{s}]=\mathbb{E}[x]\\&=\int_{0}^{1}x\cdot f_X(x)dx=\int_{0}^{1}x\cdot Sf(x)(F(x))^{S-1}dx\\&=\int_{0}^{1}x\cdot \frac{Sx^{a-1}(1-b)^{b-1}}{\mathbb{B}(a,b)}(\mathbb{I}_x(a,b))^{S-1}dx,\end{split}\end{sequation}where
$\mathbb{B}(\cdot)$ is the beta function and $\mathbb{I}_x(\cdot)$
is the regularized incomplete beta function. Using the series expansion
\begin{sequation}\mathbb{I}_x(a,b)=\frac{x^a}{\mathbb{B}(a,b)}\sum_{k=0}^\infty\frac{(1-b)_{k}x^k}{(a+k)k!},\nonumber\end{sequation}
\eqref{eq14} can be expressed as\begin{sequation}\begin{split}&\mathbb{E}[\bar{\gamma}_{s^*}]\\&=\frac{S}{(\mathbb{B}(a,b))^S}\int_{0}^{1}x^a(1-x)^{b-1}\left(x^a\sum_{k=0}^\infty\frac{(1-b)_kx^k}{(a+k)k!}\right)^{S-1}\\&=\frac{S}{(\mathbb{B}(a,b))^S}\\&\quad\times\sum_{k_1=0}^\infty\cdots\sum_{k_{S-1}=0}^\infty\frac{(1-b)_{k_1}\cdots (1-b)_{k_{S-1}}}{(a+k_1)\cdots (a+k_{S-1})k_1!\cdots k_{S-1}!}\\&\quad\times\int_{0}^{1}x^{aS+k_1+\cdots+k_{S-1}}(1-x)^{b-1}dx\\&=\frac{S}{(\mathbb{B}(a,b))^S}\\&\quad\times\sum_{k_1=0}^\infty\cdots\sum_{k_{S-1}=0}^\infty\frac{(1-b)_{k_1}\cdots (1-b)_{k_{S-1}}}{(a+k_1)\cdots (a+k_{S-1})k_1!\cdots k_{S-1}!}\\&\quad\times\mathbb{B}(aS+k_1+\cdots+k_{S-1}+1,b),\end{split}\nonumber\end{sequation}where
$(\cdot)_k$ is the Pochhammer symbol defined as $(x)_k=x(x+1)\cdots(x+k-1).$

\bibliographystyle{IEEEtran}
\bibliography{OIA}

\end{document}